\documentclass[a4paper,fleqn,usenatbib]{mnras}

\usepackage{newtxtext,newtxmath}

\usepackage[T1]{fontenc}
\usepackage{ae,aecompl}

\usepackage{graphicx}	
\usepackage{amsmath}	
\usepackage{amssymb}	

\title[Dust in $z>7$ galaxies]{Expected dust grain size distributions in galaxies detected by ALMA at $z>7$}

\author[H.-M. Liu and H. Hirashita]{
Hsin-Min Liu$^{1,2,3}$\thanks{E-mail: shayna501@yahoo.com.tw} and
Hiroyuki Hirashita$^1$
\\
$^{1}$Institute of Astronomy and Astrophysics, Academia Sinica,
Astronomy-Mathematics Building, AS/NTU, No.\ 1, Sec.\ 4, Roosevelt Road, Taipei 10617, Taiwan \\
$^{2}$Department of Optoelectric Physics, Chinese Culture University, No.\ 55, Hwa-Kang Rd., Yang-Ming-Shan, Taipei 11114, Taiwan\\
$^3$Institute of Astronomy, National Tsing Hua University, No.\ 101, Sec.\ 2, Kuang-Fu Road, Hsinchu 30013, Taiwan \\
}

\date{Accepted XXX. Received YYY; in original form ZZZ}

\pubyear{2019}

\begin{document}
\label{firstpage}
\pagerange{\pageref{firstpage}--\pageref{lastpage}}
\maketitle

\begin{abstract}
The dust properties in high-redshift galaxies provide clues to the origin of dust
in the Universe.
Although dust has been detected in galaxies at redshift $z>7$, it is difficult to
constrain the dominant dust sources only from the
total dust amount. Thus, we calculate the evolution of grain size distribution,
expecting that different dust sources predict different grain size distributions.
Using the star formation time-scale and the total baryonic mass constrained by
the data in the literature, we calculate the evolution of grain size distribution.
To explain the total dust masses in ALMA-detected $z>7$ galaxies,
the following two solutions are possible: (i) high dust condensation efficiency in
stellar ejecta,
and (ii) efficient accretion (dust growth by accreting the gas-phase metals in the interstellar
medium). We find that these two scenarios predict significantly different grain size
distributions: in (i), the dust is dominated by large grains ($a\gtrsim 0.1~\micron$, where
$a$ is the grain radius), while in (ii), the small-grain ($a\lesssim 0.01~\micron$)
abundance is significantly enhanced by accretion.
Accordingly, extinction curves are expected to be much steeper in (ii) than in (i). Thus, we conclude that extinction curves provide a viable way to distinguish the dominant dust sources in the early phase of galaxy evolution.
\end{abstract}


\section{Introduction}

Galaxy formation is a fundamental problem in the cosmic structure formation and is
still a mystery.
Although it is difficult to directly observe the first galaxies in the Universe,
recent advanced telescopes have enabled us to explore high redshifts.
In particular, the metals produced by the first generation of stars
provide useful tracers for detecting galaxies
through metal emission lines. Recent success in identifying
galaxies at $z\sim 9$ ($z$ is the redshift) using the [O \textsc{iii}] 88 $\micron$ line by the
Atacama Large Millimetre/submillimetre Array (ALMA) \citep{Hashimoto:2018aa}
is a good example for the current exploration of the redshift frontier.

The solid phase of metals -- dust -- has some important influences
on galaxy evolution. Dust absorbs the stellar light and reemits it in the far-infrared (FIR).
Therefore, the spectral energy distributions (SEDs) of galaxies are largely affected by
dust \citep[e.g.][]{Takeuchi:2005aa,Aoyama:2019aa}. Moreover, the surface of dust is the main site of
the formation of some molecular species, creating a molecular-rich environment which is
favorable for star formation \citep[e.g.][]{Cazaux:2004aa,Chen:2018aa}. Dust also affects the temperature of a star-forming gas through dust cooling, which induces the fragmentation in the final stage of star formation \citep[e.g.][]{Omukai:2005aa,Schneider:2006aa}.
All the above processes are influenced not only by the total dust amount but also by the total dust surface area, which is determined by the grain size distribution.
In this context, it is necessary to clarify the evolution of the grain size distribution as well as that of the total dust abundance in galaxies.

The current redshift frontier of dust observations is around $z\sim 7$--8 and is achieved
by ALMA \citep[e.g.][]{Dayal:2018aa}.
The ALMA submillimetre and millimetre bands are powerful in detecting dust emission from `normal' galaxy populations represented by Lyman break galaxies (LBGs) at $z>7$
\citep{Watson:2015aa,Laporte:2017aa,Hashimoto:2018ab,Tamura:2019aa}.
We refer to galaxies whose dust emission is detected by ALMA as ALMA-detected galaxies
in this paper.
Dust continuum together with the [O\,\textsc{iii}] 88~$\micron$ and
[C\,\textsc{ii}] 158~$\micron$ lines has become a powerful tracer for the physical conditions of galaxies found in the redshift frontier. This in turn emphasizes the importance of understanding dust itself.

To give a quantitative understanding to the dust at high redshift, it is necessary to clarify the major source of dust.
In the early stage of galaxy evolution, stellar sources control
the total dust abundance \citep[e.g.][]{Valiante:2009aa}.
Dust mass grows also through the accretion of gas-phase metals.
This process is efficient in the dense ISM. 
\citet{Mancini:2015aa} showed that the dust mass in a LBG at $z=7.5$
detected by ALMA \citep{Watson:2015aa} can be explained only if dust growth by accretion is very efficient.
\citet{Wang:2017aa} supported this conclusion, but they also found that high dust condensation efficiency combined with weak supernova (SN) dust destruction could also explain the rich dust content in the ALMA-detected high-redshift galaxies.
If we only see the total dust amount, however, it is difficult to distinguish between the above two mechanisms of dust production, namely, efficient dust condensation in stellar ejecta and quick dust growth by accretion
\citep{Lesniewska:2019aa}.

The grain size distribution is another important aspect that characterizes the dust evolution in the ISM. 
\citet{Asano:2013aa} showed that various processes driving the dust evolution put different imprint in the grain size distribution.
In particular, the dust abundance is dominated by large grains
when the stellar dust production is the major dust source, since dust grains ejected from stars are considered to be biased to large ($a\gtrsim 0.1~\micron$, where $a$ is the grain radius) grains
\citep[e.g.][]{Nozawa:2007aa,Yasuda:2012aa,DellAgli:2017aa}.
Observations of the optical properties of dust formed in stellar environments (mainly in stellar winds)
also show that dust condensed in the ejecta is dominated by large (sub-micron) grains
\citep{Groenewegen:1997aa,Winters:1997aa,Scicluna:2015aa}.
Dust growth by accretion starts to dominate the total dust abundance when the ISM is
enriched with metals (typically $Z>0.1$ Z$_{\sun}$, where $Z$ is the metallicity), and it drastically increases the small ($a\lesssim 0.01~\micron$) grain abundance.
Since the change of grain size distribution could be observable
through extinction curves \citep[e.g.][]{Hou:2016aa}, it is worth investigating the
possibility of distinguishing the dominant dust production mechanisms through the grain
size distribution.
In this paper, we aim at calculating the grain size distributions and extinction curves expected for ALMA-detected galaxies at $z>7$ to examine whether or not the different dust enrichment mechanisms could be distinguished through those quantities.

This paper is organized as follows.
We present our models for galaxy and dust evolution in Section \ref{sec:model}.
In Section \ref{sec:obs}, we explain the observational data we use. We show the results on the total dust mass and the grain size distribution in Section \ref{sec:result}. We also present our predictions for extinction curves.
We discuss our results in Section \ref{sec:discussion} and conclude in  Section \ref{sec:conclusion}.  
For the cosmological parameters, we adopt $H_0= 70$ km s$^{-1}$ Mpc$^{-1}$ , $\Omega _{\rm {M}} = 0.3$ and $\Omega_{\Lambda} = 0.7$.

\section{Model}\label{sec:model}

We use the model developed by HA19 for the evolution of grain size distribution in a galaxy.
For this model, we also need a galaxy evolution framework -- chemical evolution model
-- to calculate the enrichment
of metals, which is tightly related to dust evolution \citep{Lisenfeld:1998aa,Dwek:1998aa}.
In what follows, we
explain the galaxy evolution model, followed by a review of HA19's dust evolution
model. We model the galaxy as a one-zone object for simplicity.

\subsection{Galaxy evolution model}\label{subsec:chemical}

We adopt a simple chemical evolution model adopted by \citet{Wang:2017aa}.
We assume that the
galaxy is a closed box starting from the total baryonic mass of $M_\mathrm{g,0}$.
We also apply the instantaneous recycling approximation to simplify the calculation;
however, as shown by \citet{Hirashita:2011aa}, the metallicity evolution is not significantly
altered by this simplification as long as the galaxy age is much older than $10^7$ yr.
The gas mass, $M_\mathrm{g}$, evolves as the star formation proceeds as
\begin{align}
\frac{\mathrm{d}M_\mathrm{g}}{\mathrm{d}t}= -(1-\mathcal{R})\psi ,\label{eq:Mg}
\end{align}
where $\psi$ is the star formation rate (SFR) and $\mathcal{R}$ is the returned
fraction of the gas from
dying stars (we adopt the instantaneous recycling approximation).
We assume that the SFR is described by a constant star formation
time-scale $\tau_\mathrm{SF}$ as
 \begin{align}
\psi =M_\mathrm{gas}/ \tau_\mathrm{SF}.
 \end{align}
The evolution of the metallicity, $Z$, is written as \citep{Wang:2017aa}
\begin{align}
\frac{\mathrm{d}Z}{\mathrm{d}t}=\frac{y}{\tau_{\rm{SF}}},
\end{align}
where $y$ is the metal yield.
We also obtain the stellar mass, $M_\star$ as
\begin{equation}
M_\star = M_\mathrm{g,0} - M_\mathrm{g}.
\end{equation}
We adopt $M_\mathrm{g}=M_\mathrm{g,0}$ and $Z=0$ at $t=0$ for the initial condition,
and $\mathcal{R}=0.16$ and $y=0.014$ following \citet{Wang:2017aa}.
We constrain $M_\mathrm{g,0}$ and $\tau_\mathrm{SF}$ using the
stellar masses, SFRs, and ages
derived from the observational data described in Section \ref{sec:obs}.
After fixing these two quantities,
we obtain $Z(t)$, which is used as an input for the dust evolution calculated in
the next subsection.

\subsection{Evolution of grain size distribution}\label{subsec:dustmodel}

The evolution of grain size distribution has been formulated in our previous paper
(HA19) based on \citet{Asano:2013aa}.
We avoid repeating the description for the formulation, and only mention the essence and
important parameters. We refer the interested reader to HA19 for the full explanation.

We assume grains to be spherical and compact,
so that the mass of a grain is written as $m=(4\upi /3)a^3s$, where $a$ is the grain radius
and $s$ is the material density of dust.
We adopt $s=3.5$ g cm$^{-3}$ based on silicate in
\citet{Weingartner:2001aa} for the calculation of grain size distribution.
We adopt the same parameter values as
in the one-zone galaxy model of HA19 unless otherwise stated.

The grain size distribution at time $t$, $n(a,\, t)$, is defined such that
$n(a,\, t)\,\mathrm{d}a$ is the number density
of dust grains whose radius is between $a$ and $a+\mathrm{d}a$.
The total dust mass density $\rho_\mathrm{d,tot}(t)$ is calculated by
\begin{align}
\rho_\mathrm{d,tot}(t)=\int_0^\infty\frac{4}{3}\upi a^3s\, n(a,\, t)\,\mathrm{d}a.
\end{align}
Since the gas density is given by the number density of hydrogen nuclei,
$\rho_\mathrm{gas}=\mu m_\mathrm{H}n_\mathrm{H}$
($\mu =1.4$ is the gas mass per hydrogen, $n_\mathrm{H}$, given later, is the
number density of hydrogen nuclei, and $m_\mathrm{H}$ is the mass
of hydrogen atom), the dust-to-gas ratio, $\mathcal{D}(t)$, is estimated
as
$\mathcal{D}(t)={\rho_\mathrm{d,tot}(t)}/{\rho_\mathrm{gas}}$.
Using this, the total dust mass, $M_\mathrm{dust}(t)$ is estimated as
\begin{align}
    M_\mathrm{dust}(t)=\mathcal{D}(t)M_\mathrm{g}(t).
\end{align}

The time evolution of the grain size distribution is driven by the
following processes: dust condensation in stellar ejecta, dust destruction by SNe,
grain disruption by shattering, dust growth by the accretion of gas-phase metals,
and grain growth by coagulation.
We assume that shattering only occurs in the
diffuse ISM (occupying half of the ISM) and that accretion and coagulation take
place in the dense ISM (occupying the other half; we denote the dense gas fraction
as $f_\mathrm{dense}$, which is 0.5 here).
The other processes occur in both phases.
The gas density and temperature are
$(n_\mathrm{H},\, T_\mathrm{gas})=(0.3~\mathrm{cm}^{-3},\, 10^4~\mathrm{K})$,
and $(300~\mathrm{cm}^{-3},\, 25~\mathrm{K})$
in the diffuse and dense ISM, respectively.
The variations of $n_\mathrm{H}$ and $T$ are degenerate with that of $f_\mathrm{dense}$;
the change of $f_\mathrm{dense}$ indeed affects
the dust evolution, but it does not influence our conclusions as long as
we adopt a model that explains the dust mass at the relevant ages.
Since $f_\mathrm{dense}$ is not the dominant factor for the conclusions drawn in this paper,
we fix it to 0.5.

In this paper, although we include all the above dust evolution processes,
we focus on stellar dust production, SN dust destruction, and dust growth by accretion.
This is because these three processes have the largest impacts on
both total dust abundance and grain size distribution (HA19).
The parameters regarding the efficiencies of those processes (described below) are constrained by ALMA data.
We fix the parameters concerning the other processes, shattering and coagulation to those adopted in HA19.

The increase of dust mass by stellar dust production is calculated by
assuming a constant
dust condensation efficiency $f_\mathrm{in}$.
The increasing rate of the total dust mass density is thus estimated as
\begin{align}
\dot{\rho}_\mathrm{d,tot}=f_\mathrm{in}\rho_\mathrm{gas}\dot{Z},
\end{align}
where $\dot{Z}\equiv\mathrm{d}Z/\mathrm{d}t$ is evaluated in Section \ref{subsec:chemical}.
The increased dust mass density is distributed to the entire grain radius range based on
a lognormal function with a central grain radius of 0.1 $\micron$ and a standard deviation of 0.47 (see \citealt{Asano:2013ab} for the choice of these values).
The detailed functional form of the size distribution of grains formed by stellar sources does not affect our conclusions as long as it is dominated by large ($\gtrsim 0.1~\micron$) grains.
The dominance of large grains is supported by theoretical and observational studies as we mentioned in the Introduction, but we discuss it further in Section \ref{subsec:robustness}.
There is an uncertainty in $f_\mathrm{in}$ as seen in
in different dust condensation calculations \citep{Inoue:2011aa,Kuo:2013aa},
so that 
we investigate a range of $f_\mathrm{in}=0.01$--0.5 with a fiducial value of $f_\mathrm{in}=0.1$.

The SN destruction is regulated by the destruction time-scale, $\tau_\mathrm{dest}$,
evaluated as a function of grain mass by
\begin{align}
\tau_\mathrm{dest}(m)= \frac{M_\mathrm{g}}{\epsilon_\mathrm{dest}(m)M_\mathrm{SN}\gamma},
\end{align}
where $M_{\rm{SN}}$ is the gas mass swept by a single SN, $\gamma$ is the SN rate,
and  $\epsilon_\mathrm{dest}(m)$ is the dust destruction efficiency as a function of grain mass $m$ \citep{McKee:1989aa}.
Note that we use the gas mass $M_{\rm{g}}$ calculated by equation (\ref{eq:Mg}).
The destruction efficiency is a decreasing function of the grain mass (radius) with $\epsilon_\mathrm{dest}= 0.1$ at $a= 0.1~\micron$. 
To investigate various SN destruction efficiencies, we change $M_\mathrm{SN}$.
A high/low value of $M_{\rm{SN}}$ enables us
to examine cases with strong/weak (or efficient/inefficient) SN destruction. We set the fiducial value as $M_\mathrm{SN}=6.8 \times 10^3$ M$_\odot$ (HA19),
and investigate a range of $6.8$--$6.8 \times 10^{4}$ M$_{\sun}$.

For dust growth by accretion, we use the grain-radius-dependent accretion time-scale as
\begin{align}
\tau_\mathrm{acc}(m) &= \frac{1}{3} \tau_\mathrm{0,acc}\left(\frac{a}{0.1~\micron}
\right)\left(\frac{Z}{\mathrm{Z}_{\sun}}\right)^{-1},
\label{eq:tau_acc}
\end{align}
where $\tau_\mathrm{0,acc}$ is a constant parameter.
We adopt $\tau_\mathrm{0,acc}=1.6\times 10^8$ yr
for the fiducial case based on the value for silicate \citep{Hirashita:2012aa}.
This value
is sensitive to the gas density and temperature.
In order to examine various dust growth efficiencies, we change $\tau_\mathrm{0,acc}$ in the range from $5.4 \times 10^6$ to  $1.6 \times 10^8$ yr.
The most important feature in equation (\ref{eq:tau_acc}) is that smaller grains have shorter accretion time-scales.
This means that the effect of accretion appears more prominently at smaller grain sizes.

As mentioned in Section \ref{subsec:chemical}, there is no dust initially. The grain size distribution follows the lognormal function adopted for the stellar dust production in the earliest epoch, but is subsequently modified by the interstellar processing.
The fiducial values and ranges of the parameters are summarized in Table \ref{tab:param}.

\begin{table}
\tabcolsep=2pt
\caption{Fiducial values and ranges of
the parameters.}
\begin{center}
    \begin{tabular}{lcccc}
     \hline
     Process & Parameter & Fiducial value & Minimum & Maximum \\
     \hline
     Stellar dust & $f_{\rm{in}}$ & 0.1 & 0.1 & 1 \\
     Sputtering & $M_{\rm{SN}}$ & $6.8\times 10^3$ M$_\odot$ & 68 M$_\odot$ & $6.8 \times 10^4$ $ \rm {M_\odot}$\\
    Accretion & $\tau_\mathrm{acc}$ & $1.61 \times 10^8$ yr &  $5.4 \times 10^6$ yr & $1.61 \times 10^8$ yr \\
    \hline
   \end{tabular}
\label{tab:param}
\end{center}
\end{table}

\subsection{Extinction curves}\label{subsec:extinc}

As an observable signature that could constrain the grain size distribution,
we calculate the extinction curve based on the obtained grain size distribution.
Although it is known that the same extinction curve could be reproduced by different
grain size distributions with different grain materials \citep{Zubko:2004aa},
we still expect that we are able to discriminate between some
extreme grain size distributions \citep{Hou:2016aa}. Therefore, as a first step, we make an attempt to predict grain
size distribution assuming often used grain properties for nearby galaxies.

The extinction curve (extinction as a function of wavelength, $A_\lambda$)
is calculated by the following equation:
\begin{align}
A_{\lambda}=(2.5 \log_{10}\mathrm{e})L\displaystyle\sum_{i}\displaystyle\int_{0}^{\infty}
\pi a^{2}Q_{\rm ext}(a, \lambda)n_i(a)\,\mathrm{d}a,
\end{align}
where $Q_{\rm ext}(a,\,\lambda)$ is the extinction cross-section
normalized to the geometrical cross-section,
$n_i(a)$ is the grain size distribution of the $i$th dust component
(see below), and
$L$ is the path length.
Although our dust evolution model is not capable of separating the grain
species, we assume a mixture of two components (silicate and carbonaceous dust) in the calculation of extinction curves \citep{Draine:1984aa}.
The extinction cross-sections are evaluated
by using the Mie theory \citep{Bohren:1983aa}
with the same optical constants
for silicate and graphite as in
\citet{Weingartner:2001aa}.
We also calculate extinction curves using amorphous carbon taken from
\citet{Zubko:1996aa} (their ACAR)
instead of graphite as a representative case without the prominent 2175~\AA\ bump.
Indeed, some studies adopted amorphous carbon to explain the bumpless extinction curves observed in various objects
\citep{Nozawa:2015aa,Hou:2016aa}.
We assume a mixture of silicate and carbonaceous dust in the calculation of extinction curves with
a number ratio of 0.43 : 0.57
\citep{Hirashita:2009ab}.
We assume that both components have the same grain size distribution.
Because this fraction is valid for the Milky Way, it is assured that the extinction curve approaches
the Milky Way extinction curve if the \citet{Mathis:1977aa} (the so-called MRN) grain size distribution is realized for the silicate--graphite
mixture.
Since we are interested in the extinction curve shape, we output $A_\lambda /A_V$
so that $L$ is canceled out.

\section{Observational data}\label{sec:obs}
In order to constrain the earliest dust enrichment in the Universe,
we compile the data of `normal'\footnote{This indicates that we excluded
extremely bright objects such as quasars in order to trace a normal population of galaxies.}
ALMA-detected galaxies (LBGs) at $z>7$.
(See \citet{Mancini:2016aa} for a statistical approach.)
As explained in Section \ref{subsec:chemical}, the initial gas mass ($M_\mathrm{g,0}$)
and the star formation time-scale ($\tau_\mathrm{SF}$) are necessary
to fix the chemical evolution model. However, these quantities are not directly observable.
Thus, we calculate two observable quantities, the SFR and the stellar mass, to constrain
$\tau_\mathrm{SF}$ and $M_\mathrm{g,0}$. We summarize the selected galaxies
in Table \ref{tab:sample}.

\begin{table}
\renewcommand{\arraystretch}{1.2}
\tabcolsep=2pt
\centering
\begin{minipage}{80mm}
\caption{High redshift $(z>7)$ galaxies whose dust emission is detected by ALMA.}
\label{tab:sample}
    \begin{tabular}{lcccccc}
     \hline
     Name & $z$ & SFR & $M_\star$ & $M_\mathrm{dust}$ & Age &
     Ref.\,$^\mathrm{a}$\\
      & & (M$_\odot$ yr$^{-1}$) & ($10^9$ M$_\odot$) & ($10^7$M$_\odot$) & (Gyr) \\
     \hline
     MACS0416-Y1$^\mathrm{b}$ & 8.3 & 30 & $3$ & $8.2 \pm 1.6$ & $0.3$ & 1\\
     MACS0416-Y1$^\mathrm{b}$ & 8.3 & 30 & $3$ & $3.6 \pm 0.7$ & $0.3$ & 1\\
     A1689-zD1 & 7.5 & $12^{+4}_{-2}$ & $1.7^{+0.7}_{-0.5}$ & $4 ^{+4}_{-2}$ & 
     $0.081^{+0.07}_{-0.03}$ & 2\\
     A2744-YD4 & 8.38 & $20.4^{+17.6}_{-9.5}$ & $1.97^{+1.45}_{-0.66}$ & $0.55 ^{+1.96}_{-0.17}$ & $<0.607^\mathrm{c}$ & 3\\
     B14-65666$^\mathrm{d}$ & $7.15$ & $10.5 \pm 2.1$ & $2.1^{+1.1}_{-1.4} $& $5.6 \pm 1.1$& $0.2$ & 4\\
     B14-65666$^\mathrm{d}$ & $7.15$ &$ 10.5 \pm 2.1$ & $2.1^{+1.1}_{-1.4} $& $2.2 \pm 0.4$& $0.2$ & 4\\
     B14-65666$^\mathrm{d}$ & $7.15$ &$ 10.5 \pm 2.1$ & $2.1^{+1.1}_{-1.4} $& $1.2 \pm 0.2$ & $0.2$ & 4\\
     \hline
    \end{tabular}

    \medskip

$^\mathrm{a}$References: 1) \citet{Tamura:2019aa};
2) \cite{Watson:2015aa}; 3) \citet{Laporte:2017aa};
4) \citet{Hashimoto:2018ab}.\\
$^\mathrm{b}$Same object but different dust temperatures
($T_\mathrm{dust}=40$ and 50 K for the first and second lines, respectively). We adopt the age, SFR, and stellar mass of the older stellar
population whose existence is permitted in the SED fitting.\\
$^\mathrm{c}$Since the age is not given in the literature, the cosmic age is put as an upper limit.\\
$^\mathrm{d}$Same object but different dust temperatures
($T_\mathrm{dust}=30$, 40, and 50 K for the first, second, and third rows, respectively).
\end{minipage}
\end{table}

We adopt the stellar mass, age, and SFR derived from the SED fitting to the rest-frame
UV--optical data.
For A2744-YD4, the age is not given in the literature, so that the cosmic age is used as an upper limit. 
For MACS0416-Y1, \citet{Tamura:2019aa} derived a young age of $\sim 3$ Myr, which is too
short to produce dust. As they mentioned, it is likely that the dust is contributed from
an older stellar population, whose existence dost not conflict with the observed SED.
Thus, we list the quantities derived for this older component by \citet{Tamura:2019aa}.
The dust mass is estimated from the ALMA bands.
For MACS0416-Y1, we show the two dust masses corresponding to different dust
temperatures (40 and 50 K) following \citet{Tamura:2019aa}, and for
B14-65666, we list the the three cases for dust temperatures 30--50 K following
\citet{Hashimoto:2018ab}. These multiple values are useful
to present the uncertainty in the dust mass on the observational side.

\section{Results}\label{sec:result}




\subsection{Stellar mass and SFR}

\begin{figure*}
\begin{center}
\includegraphics[width=0.45\textwidth]{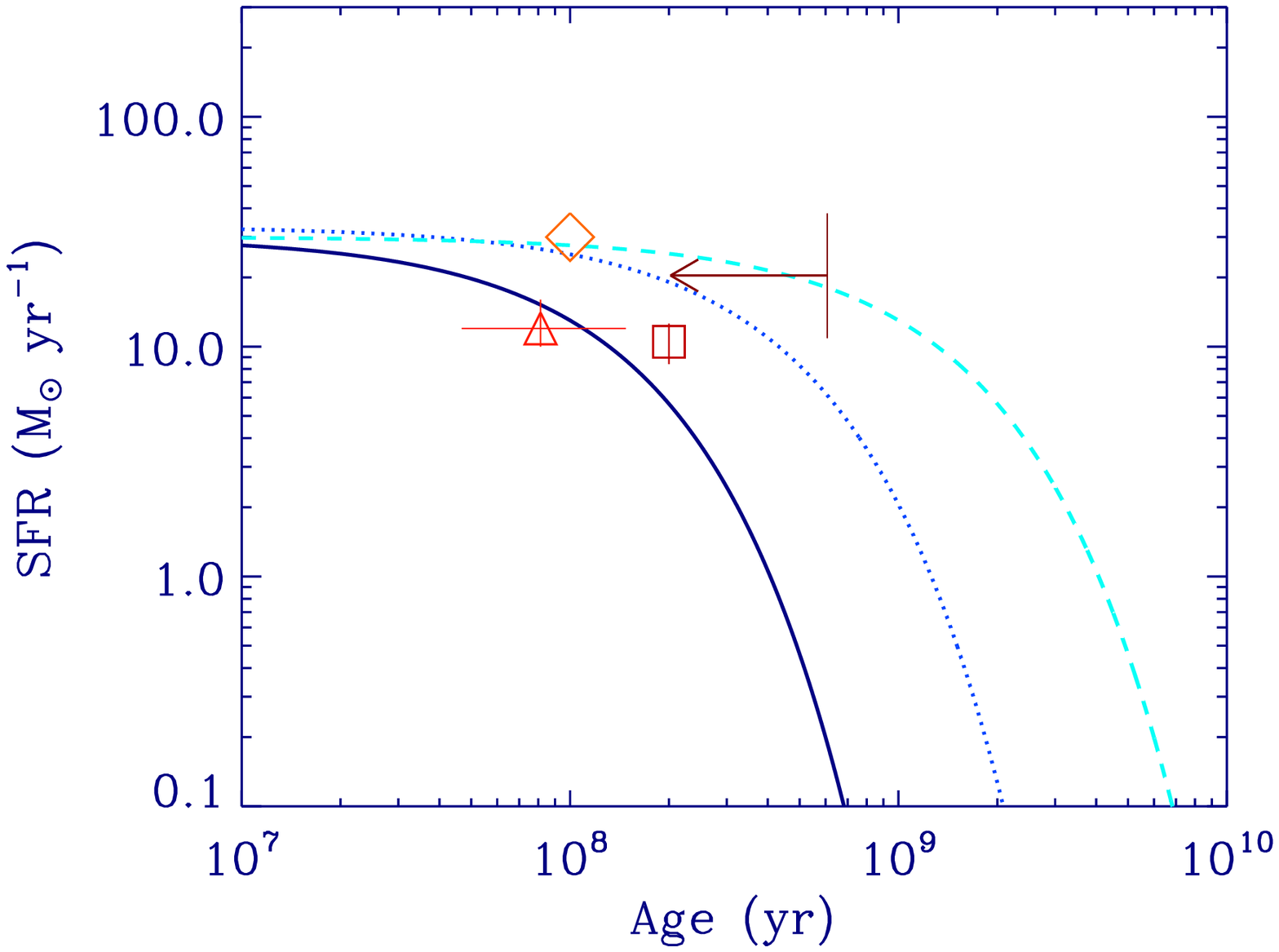}
\includegraphics[width=0.45\textwidth]{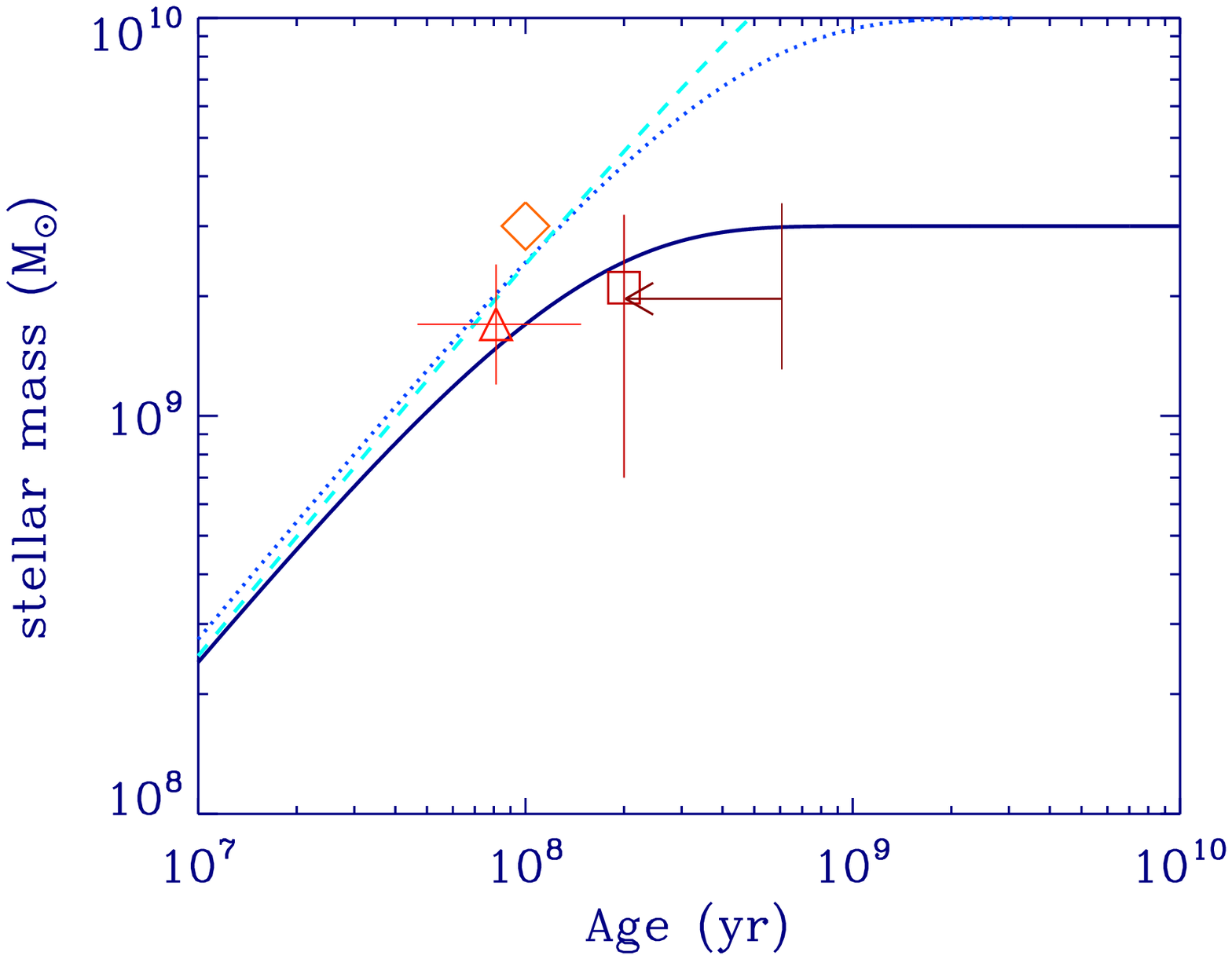}
\end{center}
\caption{Evolution of SFR (left) and stellar mass (right)
for various values of $\tau_\mathrm{SF}$ and $M_{g,0}$.
The solid, dotted, and dashed lines show Models A, B, and C, respectively
(see Table \ref{tab:param_gal} for the adopted values).
The diamond, triangle, and square show MACS0416-Y1, A1689-zD1, and B14-65666, respectively.
The arrow shows an upper limit for the age of A2744-YD4 with the vertical line
showing the range of SFR or stellar mass.
}
\label{fig:SFR}
\end{figure*}

In the model, the initial gas mass, $M_\mathrm{g,0}$,
and the star formation time-scale, $\tau_\mathrm{SF}$,
are the given parameters.
In principle, these data can be fixed using the above observational quantities
(Table \ref{tab:sample}).
The initial gas mass $M_\mathrm{g,0}$ should be larger than the observed stellar mass.
Thus, $M_\mathrm{g,0}\gtrsim 10^{10}$~M$_{\sun}$. On the other hand, to realize
SFR $\sim 10$--100 M$_{\sun}$ yr$^{-1}$, the star formation time-scale should be shorter than $\sim 10^9$ yr; otherwise, the SFR would be too low. 
Those two parameters are degenerate in the sense that a larger
$M_\mathrm{g,0}$ produces the same SFR at a certain age with a longer $\tau_\mathrm{SF}$.
Therefore, we examine the three cases for $(M_\mathrm{g,0},\,\tau_\mathrm{SF})$
as summarized in Table \ref{tab:param_gal}. The three choices are referred to as
Models A, B, and C.

\begin{table}
\tabcolsep=2pt
\caption{Parameter values adopted for the galaxy model.}
\begin{center}
    \begin{tabular}{lccc}
     \hline
     Parameters & Model A & Model B & Model C\\
     \hline
     $\tau_{\rm{SF}}$ (yr) & $3 \times 10^8$ & $10^8 $& $10^9$ \\
    $M_{\rm{g.0}}$ (M$_\odot$) & $10^{10}$ & $3 \times 10^9$ &  $3 \times 10^{10}$ \\
    \hline
   \end{tabular}
\label{tab:param_gal}
\end{center}
\end{table}

In Fig.~\ref{fig:SFR}, we show the SFR and stellar mass
as a function of age for the three models.
From the comparisons with the observational data points, we conclude that the galaxies
detected by ALMA at $z>7$ are broadly fitted with the three models.
Although the values of $M_\mathrm{g,0}$ and $\tau_\mathrm{SF}$ are uncertain
and likely to be different from galaxy to galaxy,
we expect that the three models broadly cover typical ALMA-detected cases at $z>7$.
Since, as we discuss later in Section \ref{subsec:robustness_gal}, the detailed
choice of these values do not affect
our conclusions, we adopt Model A unless otherwise stated.
We discuss Models B and C in Section \ref{subsec:robustness_gal}.

\subsection{Dust mass}\label{subsec:Mdust}
We calculate the time evolution of the dust mass.
As mentioned in Section \ref{subsec:dustmodel}, we
particularly focus on the effects of
dust formation and destruction mechanisms that directly affect the dust mass: stellar dust production, dust growth by accretion, and dust destruction by SNe,
although shattering and coagulation are also included in the calculations.
For this purpose, we change $f_\mathrm{in}$, $\tau_\mathrm{0,acc}$, and $M_\mathrm{SN}$
as described in Section \ref{subsec:dustmodel}.
Below we change one of these parameters with the others fixed to the fiducial values
(Table \ref{tab:param}).

\subsubsection{Effect of stellar dust production}
\begin{figure}
\begin{center}
\includegraphics[width=0.45\textwidth]{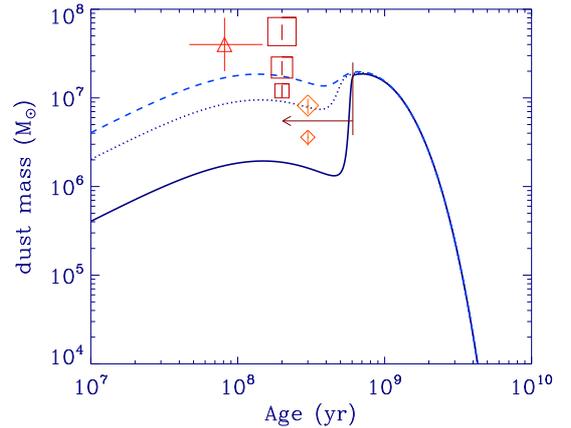}
\end{center}
\caption{Evolution of the total dust mass for various dust condensation efficiencies, 
$f_\mathrm{in}$, in Model A.
The solid, dotted, and dashed lines show the results for $f_\mathrm{in}= 0.1$, 0.5, and 1,
respectively. The observation data are shown by the same symbols as in Fig.\ \ref{fig:SFR}.
As noted in Table \ref{tab:sample}, two points for MACS0416-Y1 and three points for B14-65666 are shown for the dust masses derived for different dust temperatures (distinguished by different symbol sizes).}
\label{fig:fin}
\end{figure}

We show the evolution of dust mass for various $f_\mathrm{in}$ in Fig.~\ref{fig:fin}.
As expected, the dust mass is proportional to $f_\mathrm{in}$ at $t\lesssim 3\times 10^8$ yr.
All the cases converge to a single line at $t\gtrsim 1\times 10^9$ yr, when
the total dust mass is governed by accretion (which is not related to $f_\mathrm{in}$).
The rapid rise around $t\gtrsim 4\times 10^8$ yr caused by accretion.
This increase is more prominent for smaller $f_\mathrm{in}$ because more metals are
in the gas phase (thus they are available for dust growth).
At $t\gtrsim 5\times 10^8$~yr, the dust mass decreases because of astration
(consumption of dust and gas by star formation).

We find that $f_\mathrm{in}\gtrsim 0.5$ can explain most of the observational data.
A1689-zD1 may be marginally explained with $f_\mathrm{in}\sim 1$ considering the error bar.
We may also fail to explain B14-6566 if the dust temperature is as low as 30 K, but
the dust masses derived for 40--50 K can be explained by high dust condensation efficiencies.
Therefore, high dust condensation efficiencies ($f_\mathrm{in}\sim 0.5$--1) can be
regarded as a possible explanation of dust abundance in ALMA-detected galaxies at
high redshift.

\subsubsection{Effect of dust growth by accretion}
\begin{figure}
\begin{center}
\includegraphics[width=0.45\textwidth]{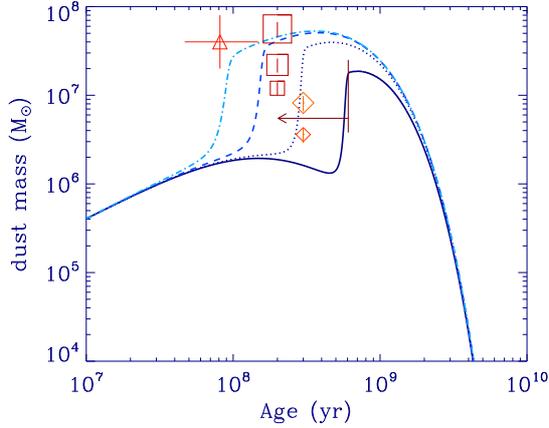}
\end{center}
\caption{Same as Fig.~\ref{fig:fin} but for various $\tau_\mathrm{0,acc}$.
The solid, dotted, dashed, and dot--dashed lines show the cases with
$\tau_\mathrm{0,acc}= 1.6 \times 10^8$, $5.4 \times 10^7$, $1.6\times 10^7$ and
$5.4\times 10^6$~yr,
respectively.
}
\label{fig:tau_acc}
\end{figure}

We change the efficiency of dust growth by accretion
in Fig.~\ref{fig:tau_acc}. We observe that efficient dust growth by accretion
($\tau_\mathrm{0,acc}\lesssim 1.6\times 10^7$ yr; that is, $\lesssim$1/10 times
the fiducial value)
explains the observational data.
If $\tau_\mathrm{0,acc}$ is shorter, the rapid increase of dust mass appears at a younger age.
The diversity in the dust mass caused by the different $\tau_\mathrm{acc}$ appears between 
$t\sim 10^8$~yr and $\sim 10^9$~yr.
All the cases with the different $\tau_\mathrm{0,acc}$ converge to the same dust mass at
$t\gtrsim 2\times 10^9$ yr, since accretion is saturated.
At $t\gtrsim 4 \times 10^8$ yr, the dust mass decreases by astration.

\subsubsection{Effect of SN destruction}\label{sput_eff}
\begin{figure}
\includegraphics[width=0.45\textwidth]{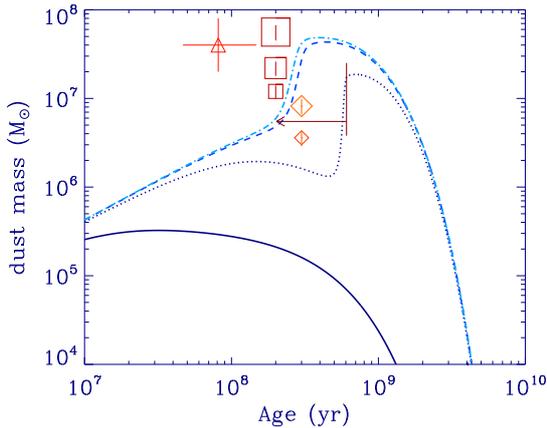}
\caption{Same as Fig.~\ref{fig:fin} but for various dust destruction efficiencies
regulated by $M_\mathrm{SN}$.
The solid, dotted, dashed, and dot--dashed lines show the result for
$M_\mathrm{SN}=6.8 \times 10^4$, $6.8\times 10^3$, $6.8\times 10^2$,
and 68 M$_{\sun}$, respectively.
}
\label{fig:SN}
\end{figure}

The dust destruction efficiency is controlled by the gas mass swept by a SN ($M_\mathrm{SN}$)
in our model. We show the evolution of dust mass for various values of $M_\mathrm{SN}$
in Fig.~\ref{fig:SN}. We observe that the change of SN destruction efficiency affects the
dust mass at the relevant ages for the sample. We also find that, even if we weaken the
SN destruction, we fail to explain the data points with high dust masses. Therefore, either
a high condensation efficiency in stellar ejecta or a short dust growth time-scale is
necessary to explain the dust masses of the sample galaxies.
On the other hand, it is also true that efficient destruction suppresses the dust mass
significantly; thus, we also conclude that dust destruction efficiency should be
similar to or weaker than that expected from the fiducial case.

\subsection{Grain size distribution}\label{subsec:size}

In the above, we have shown that there are two `solutions' to explain the dust masses
in the sample galaxies:
a high dust condensation efficiency in stellar ejecta,
and a short accretion time-scale.
The key parameters are $f_\mathrm{in}$ and $\tau_\mathrm{0,acc}$. These two solutions are indistinguishable only with the dust mass.
Our model is capable of predicting the grain size distribution; therefore, we here show the grain size distributions to examine how they are affected by the above two key parameters.
Since the sample galaxies have roughly ages of $\sim 3\times 10^8$ yr, we show the grain size distributions at $t=3\times 10^8$ yr.
The conclusions below are not altered by the precise value of $t$ as long as it is chosen within the range of the observationally derived ages.

\begin{figure}
\includegraphics[width=0.45\textwidth]{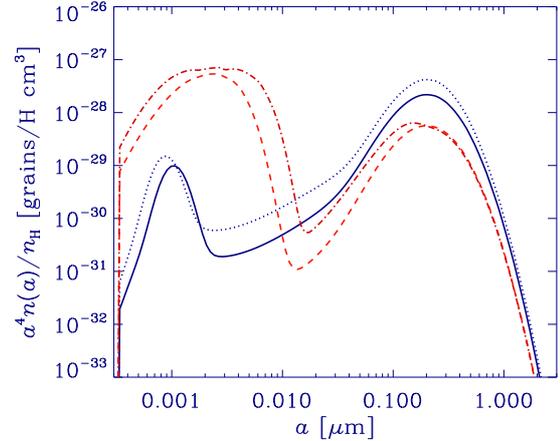}
\caption{Grain size distributions at $t=3\times 10^8$ yr for $f_\mathrm{in}=0.5$ and 1
(blue solid and dotted lines, respectively), and $\tau_\mathrm{0,acc}=5.4\times 10^8$ and
$1.61\times 10^7$ yr (red dashed and dot--dashed lines, respectively) in Model A.
The other parameters are fixed to the fiducial values (Table \ref{tab:param}).
The vertical axis shows the grain size distribution per hydrogen multiplied by $a^4$: the resulting quantity is proportional to the grain mass distribution per $\log a$.}
\label{fig:gsdfinacc}
\end{figure}

In Fig.\ \ref{fig:gsdfinacc}, we show the grain size distributions
for the two solutions.
For the dust condensation efficiency, $f_\mathrm{in}\gtrsim 0.5$ is consistent
with the observed dust masses (Fig.\ \ref{fig:fin}).
Thus, we adopt $f_\mathrm{in}=0.5$ and 1.
For dust growth, $\tau_\mathrm{0,acc}\lesssim 5.4\times 10^7$~yr is consistent with the
data points (Fig.\ \ref{fig:tau_acc}), so that we adopt $\tau_\mathrm{0,acc}=5.4\times 10^7$
and $1.6\times 10^7$~yr.
We change either $f_\mathrm{in}$ or $\tau_\mathrm{0,acc}$ and fix the other parameters
to the fiducial values (Table \ref{tab:param}).
We show the grain size distribution multipled by $a^4$, which is proportional to
the grain mass distribution per $\log a$. If the value of $a^4n(a)$ is the highest at a certain
grain radius, the dust mass is dominated by grains around that radius.

The difference in the grain size distribution is obvious between the two solutions
(high $f_\mathrm{in}$ and short
$\tau_\mathrm{0,acc}$). For high $f_\mathrm{in}$, the dust abundance is dominated by large
grains, while for short $\tau_\mathrm{0,acc}$, a drastic increase of small grains is
observed. We explain each of these two cases in what follows.

\subsubsection{Efficient dust condensation in stellar ejecta}\label{subsubsec:fin}

In Fig.~\ref{fig:gsdfinacc}, we observe that the cases with high $f_\mathrm{in}$
(= 0.5 and 1)
show grain size distributions dominated by large grains. In this case, since
the grains are predominantly produced by stars, the grain size distribution traces the
lognormal grain size distribution of dust grains condensed in the stellar ejecta.
The lower peak at a small radius is due to shattering and accretion, but this peak is much lower
than that at $a\sim 0.1~\micron$.

\subsubsection{Effecticient dust growth by accretion}
As mentioned above, grain growth by accretion drastically increases the abundance
of small grains, because they have large surface-to-volume ratio
\citep{Hirashita:2012aa,Asano:2013ab}.
Thus, if there are some small grains (produced by shattering in our case), those small grains easily accrete the gas-phase metals.
As shown in Fig.~\ref{fig:gsdfinacc}, the grain size distributions with efficient
accretion (or short $\tau_\mathrm{0,acc}$) is dominated by small ($a\lesssim 0.01~\micron$)
grains. The resulting grain size distributions are completely different from
the large-$f_\mathrm{in}$ cases shown in Section \ref{subsubsec:fin}.
Since the galaxy age is still young, coagulation
does not affect the grain size distribution. Coagulation would convert small grains
produced by shattering and accretion to large grains, making the grain size distribution
similar to the so-called MRN \citep{Mathis:1977aa} distribution (HA19).
The prominent peak at $a\lesssim 0.01~\micron$,
thus, remains to be prominent before coagulation takes place efficiently.


\subsection{Robustness against the galaxy model}\label{subsec:robustness_gal}

\begin{figure}
\begin{center}
\includegraphics[width=0.45\textwidth]{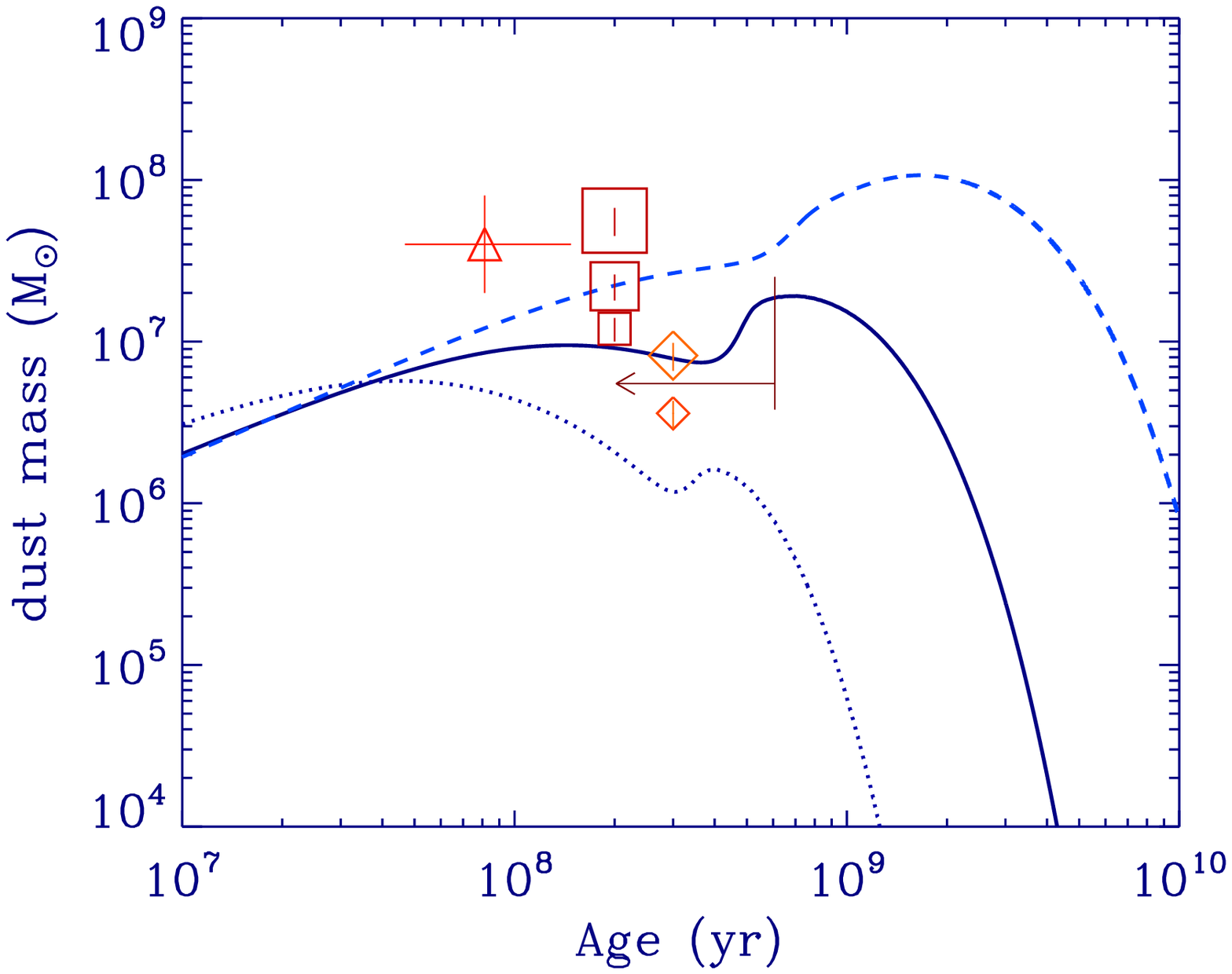}
\includegraphics[width=0.45\textwidth]{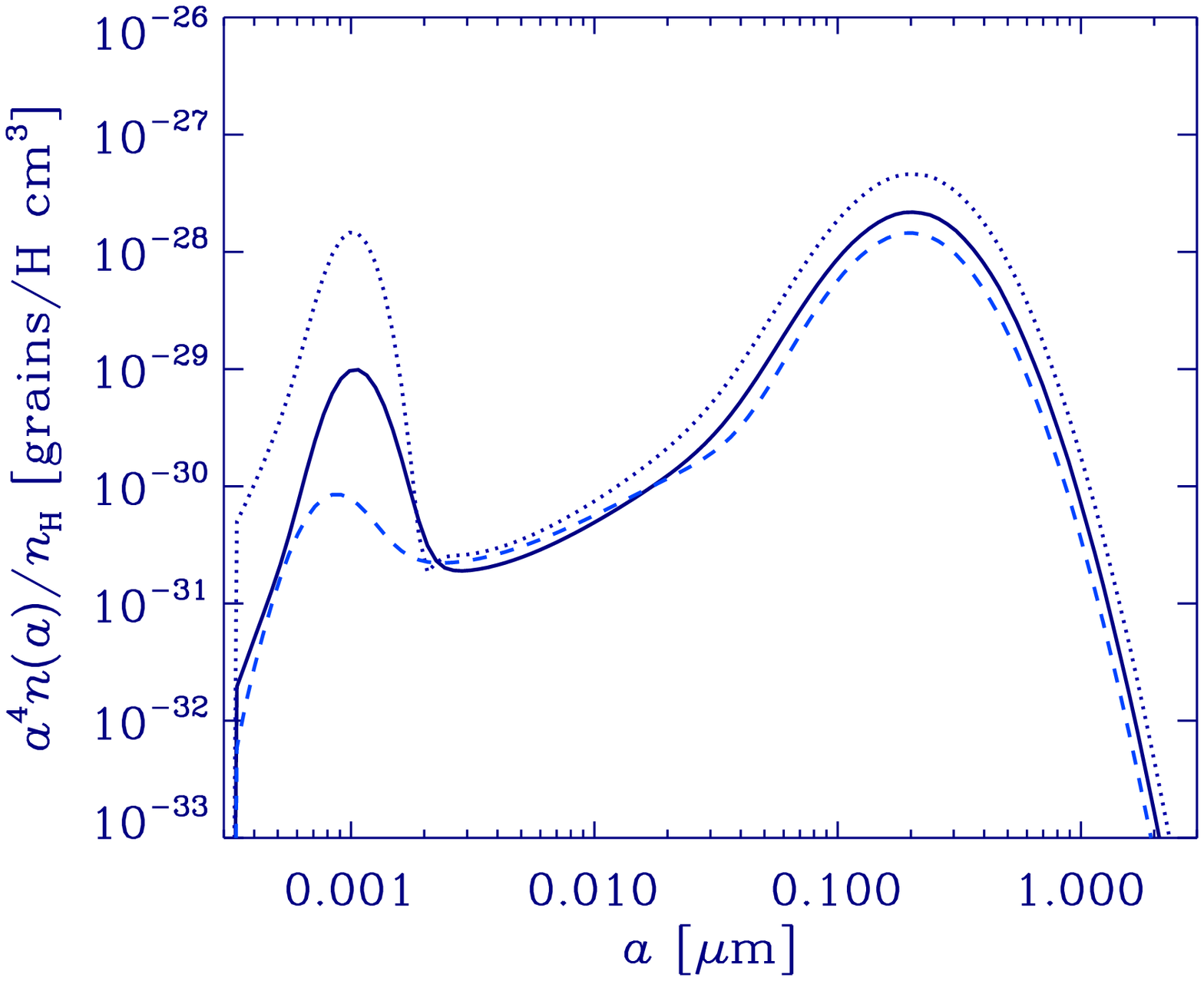}
\end{center}
\caption{Evolution of dust mass (upper) and grain size distribution (lower) for the
three galaxy models (Table \ref{tab:param_gal}). We adopt $f_\mathrm{in}=0.5$ and
the fiducial values for the other parameters (Table \ref{tab:param}).
The solid, dotted, and dashed lines show the results for
Models A, B, and C, respectively.
The same observation data as in Fig.~\ref{fig:fin} are adopted in the upper panel.
}
\label{fig:dm3fin}
\end{figure}

We have focused on Model A for the galaxy model.
Here we adopt Models B and C (Table \ref{tab:param_gal}) to examine if
the above results for Model~A robustly hold or not.
In Fig.~\ref{fig:dm3fin}, we show the evolution of dust mass and the grain size
distribution at $t=3\times 10^8$~yr for all the galaxy models. We adopt
$f_\mathrm{in}=0.5$ and fix the other parameters to the fiducial values.
We observe that Model B cannot explain the observed dust masses because of its
short $\tau_\mathrm{SF}(=10^8~\mathrm{yr})$. With such a short $\tau_\mathrm{SF}$, the
effect of astration appears early, so that the dust mass starts to decrease even at
$t\sim 10^8$~yr. Therefore, the star formation time-scale should be
longer than $10^8$ yr.
The grain size distributions are similar between Models A and C
in the sense that the dust
is still dominated by large grains.
In Model~B, the peak of the small grains becomes the highest since the dust enrichment
proceeds the most quickly; however, this model does not explain the dust mass as shown
above. Therefore, we conclude that the dominance of large grains robustly holds
as long as we adopt a model that explains the SFR, stellar mass, and dust mass at the
same time.

\begin{figure}
\begin{center}
\includegraphics[width=0.45\textwidth]{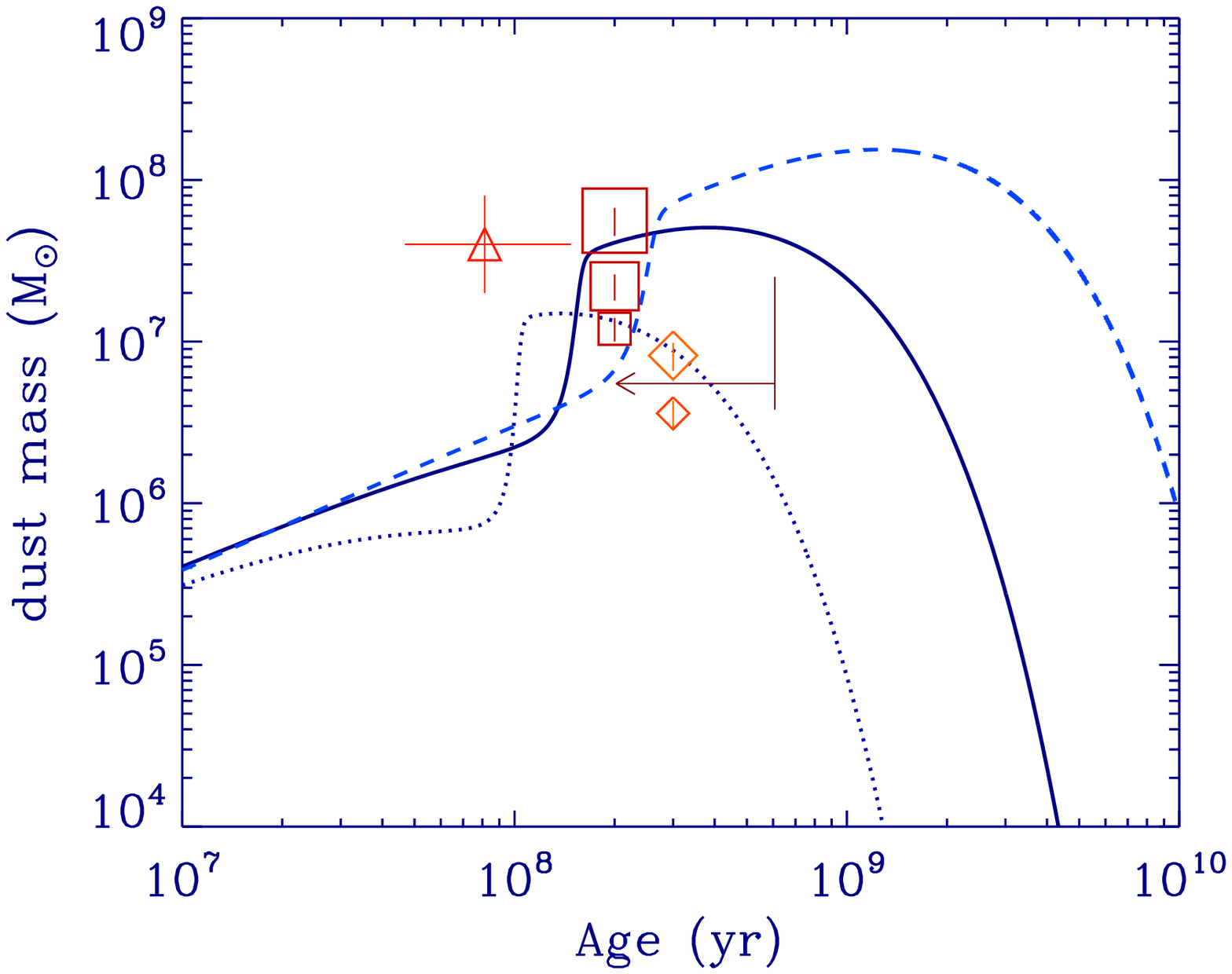}
\includegraphics[width=0.45\textwidth]{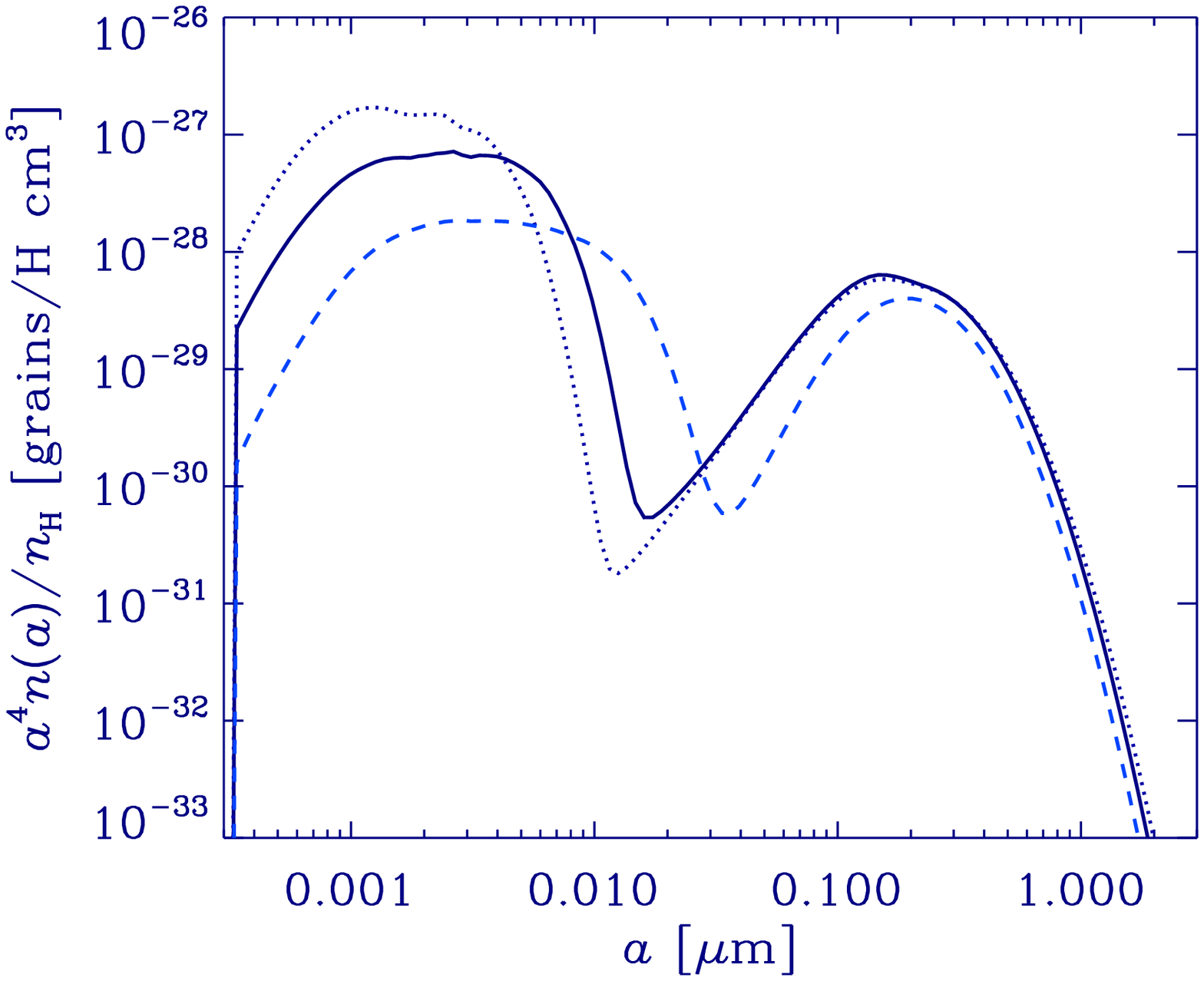}
\end{center}
\caption{Same as Fig.\ \ref{fig:dm3fin} but for $\tau_\mathrm{0,acc}=1.6\times 10^7$ yr
with the fiducial values for the other parameters (Table \ref{tab:param}).
}
\label{fig:dm3acc}
\end{figure}

In Fig.\ \ref{fig:dm3acc}, we show the results for efficient accretion with
$\tau_\mathrm{0,acc}=1.6\times 10^7$ yr for the three galaxy models.
All of the three models explain most of the data points well, although Model B shows
a little
underproduction for the dustiest objects. The lowest dust mass in Model B is due to the
most efficient astration. The grain size distributions of all the Models show
a similar small-grain-dominated feature.
Therefore, the conclusion that efficient accretion leads to strong enhancement of the grain abundance at
$a\lesssim 0.01~\micron$ is robust against the change of galaxy evolution model,
as long as the model reproduces the SFR, stellar mass, and dust mass at the same time.

The difference among the galaxy models is too small to change the extinction curves
calculated below. Thus, we
hereafter focus on Model A.

\subsection{Predicted extinction curves}\label{subsec:extinc_result}

\begin{figure}
\begin{center}
\includegraphics[width=0.45\textwidth]{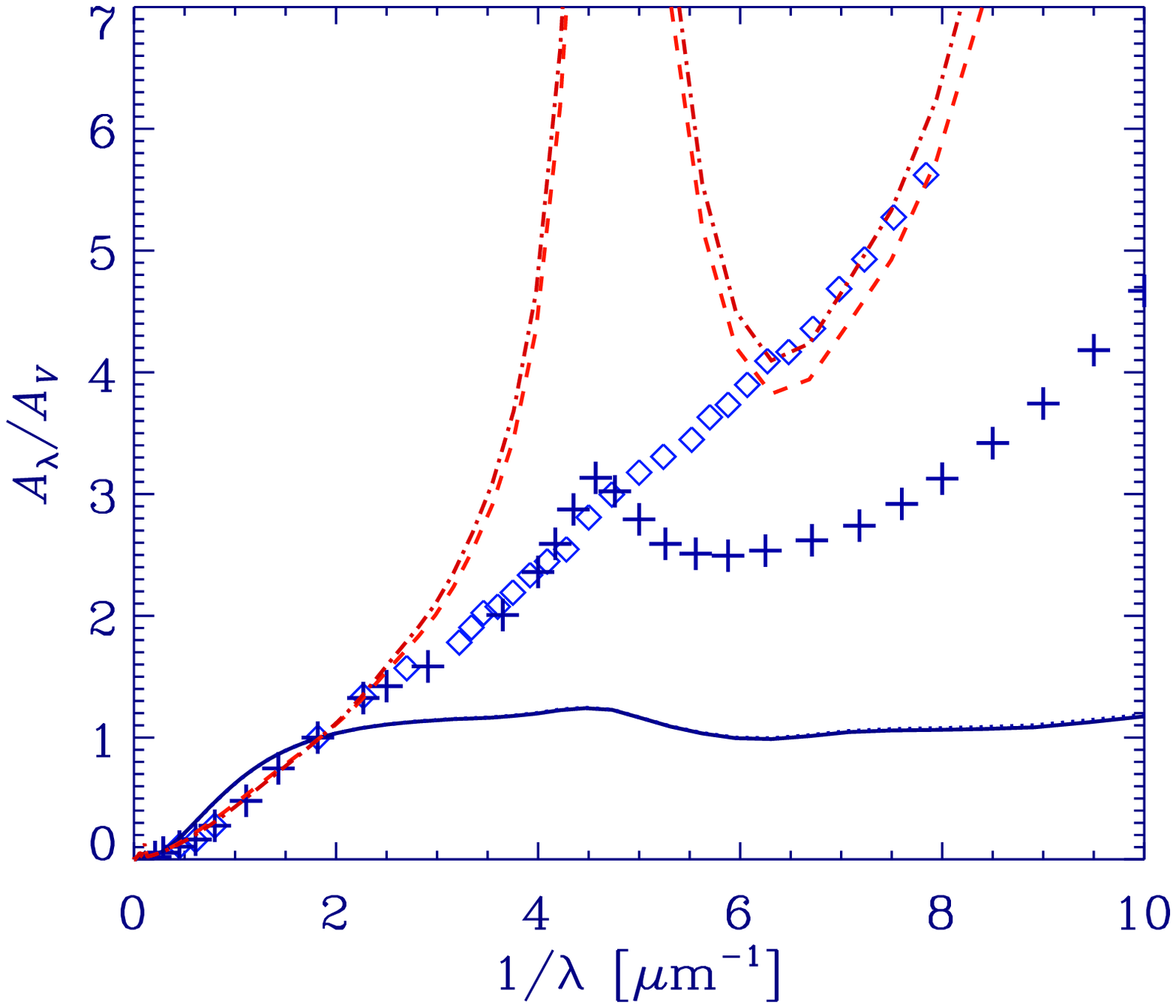}
\includegraphics[width=0.45\textwidth]{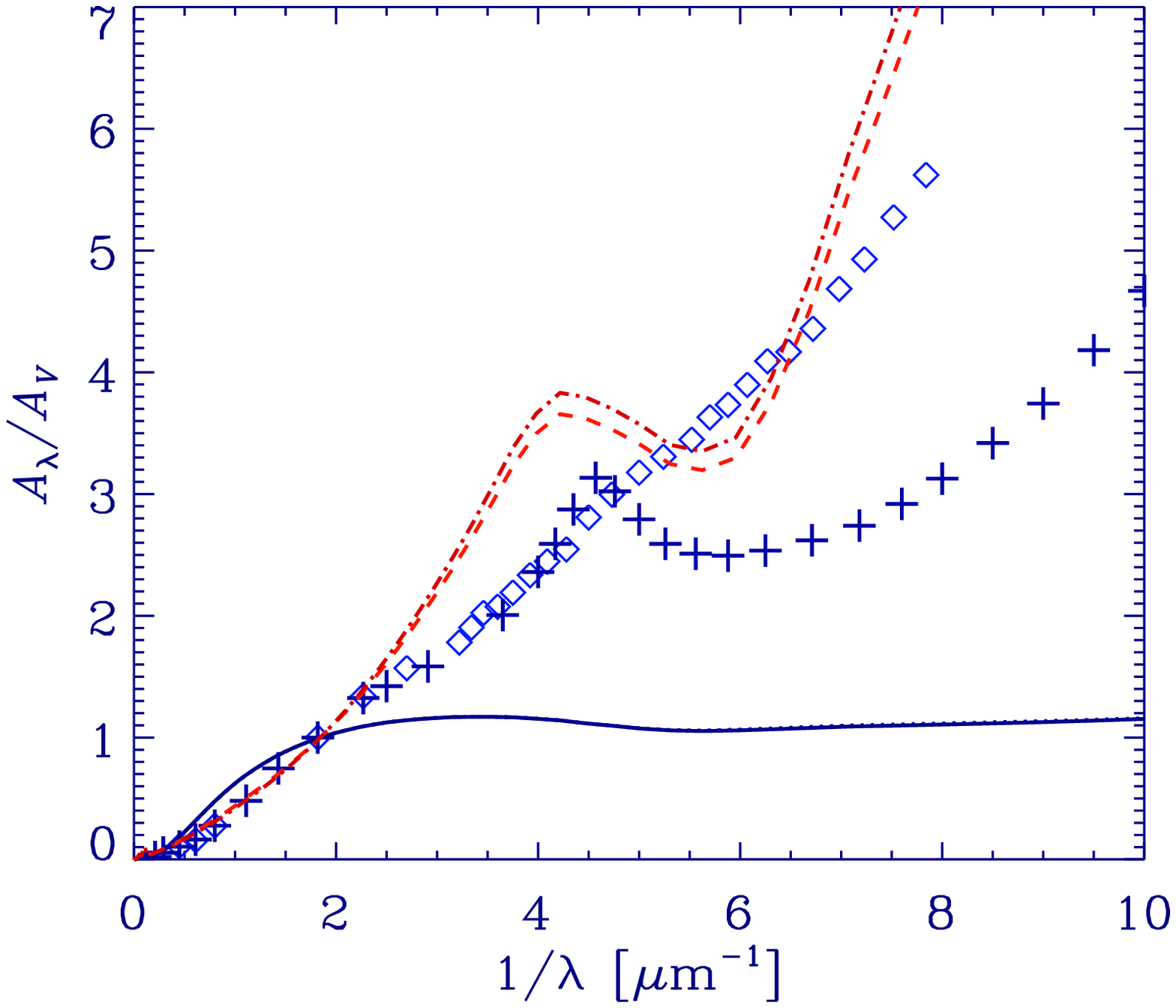}
\end{center}
\caption{Extinction curves calculated based on the grain size distributions
shown in Fig.\ \ref{fig:gsdfinacc}. The blue solid and dotted lines show the cases with
$f_\mathrm{in}=0.5$ and 1, respectively, and the red dashed and dot--dashed lines present $\tau_\mathrm{0,acc}=5.4 \times 10^8$ and $1.61 \times 10^7$ yr, respectively (the parameters
other than the varied one are fixed to the fiducial values; Table \ref{tab:param}).
The upper and lower panels correspond to the silicate--graphite and silicate--amorphous carbon mixtures, respectively.
The crosses and diamonds are the observed extinction curves
in the Milky Way and the SMC \citep{Pei:1992aa}, respectively, shown as references.}
\label{fig:ext}
\end{figure}

We calculate extinction curves based on the calculation methods describe
in Section \ref{subsec:extinc}.
In Fig.\ \ref{fig:ext}, we show the extinction curves
at $t=3\times 10^8$ yr
for the above two solutions; that is,
we chose the same cases as shown in Fig.\ \ref{fig:gsdfinacc}:
either large $f_\mathrm{in}$ (= 0.5 and 1) or short
$\tau_\mathrm{0,acc}$ ($= 5.4 \times 10^8$ and $1.61 \times 10^7$ yr).
Recall that we assume silicate--graphite and silicate--amorphous carbon mixtures.
We also show the observed extinction curves in the Milky Way and the Small Magellanic Cloud (SMC)
\citep{Pei:1992aa} just for references.

In Fig.~\ref{fig:ext}, we observe that the extinction curves for large
$f_\mathrm{in}$ are flat because the dust is dominated by large grains.
Efficient accretion ($\tau_\mathrm{0,acc}=5.4 \times 10^8$ and $1.61 \times 10^7$ yr)
produces steep extinction curves because the abundance of small grains is high.
The extinction curves in this case show an extremely prominent 2175 \AA\ bump
if we adopt graphite for the carbonaceous component,
while the extinction curve is much more smooth if we adopt amorphous carbon.
In the latter case (with amorphous carbon), although there is a small feature around wavelength
$\lambda\sim 0.25~\micron$, the extinction curves have similar slopes to the SMC extinction curve.
In summary, the extinction curves between the two solutions (efficient stellar dust production and efficient accretion) predict
drastically different extinction curves, which indicates that the two cases could be distinguished if we measure the extinction curves for high-redshift LGBs.
We further discuss this point in the next section.

\section{Discussion}\label{sec:discussion}

\subsection{Robustness of the grain sizes distributions}\label{subsec:robustness}

We adopted a grain size distribution dominated by large ($a\sim 0.1~\micron$)
grains for stellar sources based on dust condensation calculations for SN ejecta and AGB star winds (Introduction).
In particular, in the redshift range of interest ($z\gtrsim 7$), the dust is
predominantly injected by SNe rather than by AGB stars
\citep{Valiante:2009aa}. Thus, the size distribution of grains ejected from
SNe is of particular importance if the dust produced by stellar source dominates
the total dust abundance.

As shown by \citet{Nozawa:2007aa}, dust grains are
destroyed in the shocked region (by the so-called reverse shock destruction)
before being injected into the ISM. This destruction mechanism works more
efficiently for small grains for the following two reasons: (i) small grains have
large surface-to-volume ratio so that the effect of sputtering is more significant
for small grains; and (ii) small grains are more easily trapped in the shocked
region so that they suffer sputtering more.
Indeed, as shown by \citet{Nozawa:2011aa},
if SN ejecta produce only small grains, the dust grains are destroyed before being injected
into the ISM. Therefore, if the dust production is governed by stellar
sources at
$z\gtrsim 7$, the dust abundance should be dominated by large grains.

\citet{Bianchi:2007aa} also calculated reverse shock destruction based on \citet{Todini:2001aa}'s dust condensation models.
However, they did not include the above trapping effect [mentioned in item (ii)]; thus, small grains still survive. Even in \citet{Bianchi:2007aa}'s models, 
the extinction curves are flatter than the SMC curve because the contribution from relatively large carbon grains ($a\sim 0.03~\micron$) flattens the extinction curve at wavelengths $<2\upi a\sim 0.2~\micron$ according to the Mie theory.
Therefore, our results of relatively flat extinction curves for stellar dust marginally holds even if we adopt \citet{Bianchi:2007aa}'s models.

If stellar sources do not supply small grains, the only way to create them is
interstellar processing. As shown above, dust growth by accretion not only increases
the total dust abundance but also make small grains dominant. This is a consequence
of smaller grains having larger surface-to-volume ratios
\citep{Hirashita:2012aa} (see also
equation \ref{eq:tau_acc}). Coagulation could make these small grains large.
In nearby galaxies, the major source of
dust is considered to be accretion \citep[e.g.][]{Draine:2009aa},
and coagulation successfully creates
large ($a\sim 0.1~\micron$) grains
(HA19).
However, coagulation only occurs after $t=1$ Gyr (HA19), which
indicates that it is difficult for coagulation to take place significantly within the
cosmic age at $z\gtrsim 7$.
Even if coagulation occurs, the grain size distribution only approaches the MRN distribution (HA19), which still predicts an extinction curve much steeper than the flat curves realized in the case of the dominant stellar dust production (i.e.\ large $f_\mathrm{in}$).
Thus, we expect that the increase of small grains and the resulting steep extinction curves
robustly appear as long as accretion is the major mechanism of
dust mass increase.

We also note that \citet{Wang:2015b,Wang:2015a} suggested the existence of a distinct
micron-sized grain population based on the
flat mid-infrared extinction curves in the Milky Way.
Such a separate grain component is difficult to form by coagulation, which predicts a continuous grain size distribution extending continuously to smaller grains.
Also, it is not clear why this component
survives against shattering.
Whether or not such micron-sized grains contribute to the extinction curves in high-redshift galaxies depends on their origin.
Thus, it is important to investigate the formation of micron-sized grains
in the context of galaxy evolution.

\subsection{Comparison with previous modeling for the dust mass}

There have been some models of dust mass evolution in ALMA-detected galaxies
at high redshift ($z\gtrsim 7$). \citet{Mancini:2015aa} showed that, in order to
explain the dust mass in A1689-zD1 at $z=7.5$, dust growth by accretion should be
$\sim 10$ times more efficient than in nearby galaxies
\citep[see also][]{Wang:2017aa}. We have confirmed
the results of these papers.
\citet{Popping:2017aa} also reached practically the same conclusions
with a thorough modeling from high to low redshifts.
A possible explanation for the efficient accretion
is a high gas density ($n_\mathrm{H}\gtrsim 10^4$ cm$^{-3}$;
\citealt{Kuo:2013aa}).

\citet{Wang:2017aa} further discussed a possibility of high condensation efficiency
in stellar ejecta,
which equally explains the dust mass in A1689-zD1. We have also confirmed this in this paper.
This possibility of high dust formation efficiency in stellar ejecta is worth
considering given that the ISM may not have a favourable condition for dust growth
by accretion \citep{Ferrara:2016aa}. However, as discussed in
\citet{Zhukovska:2016aa}, more detailed physical processes such as charge focusing
in the accreting process of the gas-phase metals are necessary to investigate.
Experiments may give us a clue to accretion processes \citep{Rouille:2014aa}.
Before we clarify if dust growth by accretion is a viable mechanism of
reproducing the dusty objects at high redshift
or not, it may be fair to treat both efficient accretion and high
condensation efficiency as equally possible mechanisms to explain the dust content
in ALMA-detected high-redshift galaxies.
The grain
size distributions predicted in this paper provide a new clue to the dominant
dust sources.

\subsection{Grain size distribution and possible clues}

In Section \ref{subsec:size}, we have proposed that the two possibilities
of explaining the dust content of ALMA-detected $z>7$ galaxies [(i) high dust condensation efficiency in
stellar ejecta and (ii)
efficient accretion] can be distinguished by the grain size distribution. The possibilities
(i) and (ii) provide grain size distributions dominated by large and small grains, respectively.
In Section \ref{subsec:extinc_result}, we have shown that the above two scenarios
predict very different extinction curves.
Therefore, we expect that extinction curves, if they are observed, would give us 
a clue to which of those two scenarios is preferred.

Since the steepness of the extinction curve is completely different between those two
cases (i) and (ii) above, the conclusion that we can discriminate them is robust,
in spite of the fact that
extinction curves also depend on grain materials \citep{Zubko:2004aa}.
As long as stars (mainly SNe) produce grains as large as $a\sim 0.1~\micron$,
the extinction curve is flat at wavelengths $\lambda\lesssim 2\upi a\sim 0.6~\micron$
according to the Mie theory. Moreover, the small grains formed by accretion typically
have sizes $a\lesssim 0.01~\micron$, which indicates that the extinction curve has
a wavelength-dependent behaviour at UV wavelengths ($\lambda\gtrsim 0.1~\micron$)
if dust growth by accretion is dominant.

Deriving extinction curves from observations is not easy, though.
If we use the emission from an entire galaxy, we are only able to derive
the attenuation curve, which includes radiation transfer effects caused by
the detailed spatial distributions of dust and stars
\citep{Calzetti:1994aa,Witt:1996aa,Inoue:2005aa,Narayanan:2018aa}.
Different dust optical depths for different ages of stellar populations also cause a diversity
in the attenuation curves with the same extinction curve
\citep{Charlot:2000aa,Mancini:2016aa,Narayanan:2018aa}.
Point sources such as quasars and gamma-ray burst (GRB) afterglows could be used to derive extinction curves minimally affected by the effects of geometry and complicated radiation transfer.
Although these point sources are effectively used to derive extinction curves up to $z\sim 5$ \citep[e.g.][]{Maiolino:2004aa,Zafar:2011aa,Liang:2009a,Liang:2010a}, quasars are rare at $z>7$ and GRBs do not necessarily occur in desired galaxies.
However, future development of observational facilities may expand the quasar and GRB samples to $z>7$.

It would be useful to observe the rest-frame mid-infrared (far infrared in the observer's frame) dust emission features that give us a clue to the dust compositions.
If we obtain a constraint on grain materials, we have an understanding of which materials
could contribute to the extinction curve.
However, there is no near future sensitive far-infrared telescope
capable of detecting dust emission from normal galaxies at $z>7$. In this respect, observing dust emission features would be more difficult than deriving extinction curves.

We expect that, at least, attenuation curves could be obtained by the James Web Space Telescope (JWST).\footnote{https://www.jwst.nasa.gov/}
With some radiation transfer calculations under reasonable distribution geometries of stars and dust
derived from JWST and ALMA, we may be able to constrain the extinction curve.
Whether or not this idea is feasible will be examined by further modeling and tests in the future work.

\section{Conclusion}\label{sec:conclusion}

We calculate the evolution of grain size distribution by modelling galaxies
at $z>7$ whose
dust emission has been detected by ALMA (referred to as the ALMA-detected galaxies). 
We use a one-zone model that calculates
the chemical enrichment as a result of star formation, for which the star formation time-scale
and the total baryonic mass are chosen to be consistent with the observed SFR and stellar
mass. The evolution of grain size distribution is calculated in a consistent manner with the
metal enrichment. Among the various processes included in the model of dust evolution,
we focus on the following
three processes that directly affect the dust abundance: dust
condensation in stellar ejecta, dust growth by the accretion of gas-phase metals in the ISM, 
and SN destruction
(note that we also include shattering and coagulation).
The dust destruction efficiency by a SN is required to be comparable to
or lower than the fiducial value, which was used for nearby galaxies.
To investigate the two dust production mechanisms (dust formation in stellar ejecta and
dust growth by accretion), we vary the dust condensation efficiency in stellar ejecta
($f_{\rm {in}}$) and the accretion time-scale at solar metallicity
($\tau_{\rm{0,acc}}$; note that the accretion time-scale is inversely proportional to the metallicity). As a consequence, we find that high
$f_{\rm{in}}$ ($\gtrsim 0.5$) or short $\tau_{\rm{0,acc}}$ ($\lesssim 1.6\times 10^7$ yr)
explains the dust masses in the ALMA-detected galaxies.
However, it is difficult to discriminate these two `solutions' (i.e.\ high $f_\mathrm{in}$
and short $\tau_\mathrm{0,acc}$) only by the total dust mass.

We further predict the grain size distributions corresponding to
the above two solutions.
As a consequence, we find that the grain size distributions are significantly different
between these two:
the grain size distributions with large $f_{\rm{in}}$ and short $\tau_\mathrm{0,acc}$
are dominated by large
($a\sim 0.1~\micron$) and small ($a\lesssim 0.01~\micron$) grains, respectively.
This statement robustly holds as long as the model reproduces the observed dust mass,
stellar mass, and SFR at the same time.
We further calculate extinction curves based on the computed grain size distributions.
We find that the above two solutions give significantly different extinction curve shapes:
large $f_{\rm{in}}$ gives extinction curves much flatter than the Milky Way extinction
curve, while short $\tau_\mathrm{0,acc}$
produces a very steep extinction curve with a slope similar to the one seen in the
SMC extinction curve. Because the extinction curves are very different, we expect that
we are able to discriminate the major dust sources (stellar dust production or
accretion) in ALMA-detected galaxies at $z>7$ once we succeed in obtaining their
extinction curves by future observations.

\section*{Acknowledgements}

We are grateful to S.-J. Rau for her
help, and the anonymous referee for useful comments.
HH thanks the Ministry of Science and Technology for support through grant
MOST 105-2112-M-001-027-MY3 and MOST 107-2923-M-001-003-MY3
(RFBR 18-52-52-006).


\bibliographystyle{mnras}
\bibliography{references}

\bsp	
\label{lastpage}
\end{document}